\documentclass[aps,prd,twocolumn,amsmath,amssymb,superscriptaddress,nofootinbib,showpacs]{revtex4-1}
\usepackage{aas_macros}
\usepackage{graphicx}
\usepackage{bm}
\usepackage{wasysym}
\usepackage{tabularx}

\usepackage[colorlinks=true]{hyperref}

\begin{document}

\title{Constraining velocity-dependent Lorentz/CPT-violations using Lunar Laser Ranging}

\author{A.~Bourgoin}
\email{adrien.bourgoin@unibo.it}
\affiliation{Dipartimento di Ingegneria Industriale, University of Bologna, Via Fontanelle 40, Forl\`i, Italy}
\affiliation{SYRTE, Observatoire de Paris, Universit\'e PSL, CNRS, Sorbonne Universit\'e, LNE, 61 avenue de l'Observatoire 75014 Paris, France}

\author{S.~Bouquillon}
\affiliation{SYRTE, Observatoire de Paris, Universit\'e PSL, CNRS, Sorbonne Universit\'e, LNE, 61 avenue de l'Observatoire 75014 Paris, France}

\author{A.~Hees}
\affiliation{SYRTE, Observatoire de Paris, Universit\'e PSL, CNRS, Sorbonne Universit\'e, LNE, 61 avenue de l'Observatoire 75014 Paris, France}

\author{C. Le~Poncin-Lafitte}
\affiliation{SYRTE, Observatoire de Paris, Universit\'e PSL, CNRS, Sorbonne Universit\'e, LNE, 61 avenue de l'Observatoire 75014 Paris, France}

\author{Q. G. Bailey}
\affiliation{Department of Physics and Astronomy, Embry-Riddle Aeronautical University, Prescott, Arizona 86301, USA}

\author{J. J. Howard}
\affiliation{Department of Physics and Astronomy, Embry-Riddle Aeronautical University, Prescott, Arizona 86301, USA}

\author{M.-C. Angonin}
\affiliation{SYRTE, Observatoire de Paris, Universit\'e PSL, CNRS, Sorbonne Universit\'e, LNE, 61 avenue de l'Observatoire 75014 Paris, France}

\author{G.~Francou}
\affiliation{SYRTE, Observatoire de Paris, Universit\'e PSL, CNRS, Sorbonne Universit\'e, LNE, 61 avenue de l'Observatoire 75014 Paris, France}

\author{J.~Chab\'e}
\affiliation{Observatoire de la C\^ote d’Azur, G\'eoazur, Universit\'e de Nice Sophia-Antipolis, Caussols, France}

\author{C.~Courde}
\affiliation{Observatoire de la C\^ote d’Azur, G\'eoazur, Universit\'e de Nice Sophia-Antipolis, Caussols, France}

\author{J.-M.~Torre}
\affiliation{Observatoire de la C\^ote d’Azur, G\'eoazur, Universit\'e de Nice Sophia-Antipolis, Caussols, France}

\date{\today}

\begin{abstract}
  The possibility for Lorentz/CPT-breaking, which is motivated by unification theories, can be systematically tested within the standard-model extension framework. In the pure gravity sector, the mass dimension 5 operators produce new Lorentz and CPT-breaking terms in the 2-body equations of motion that depend on the relative velocity of the bodies. In this Letter, we report new constraints on 15 independent SME coefficients for Lorentz/CPT-violations with mass dimension 5 using lunar laser ranging. We perform a global analysis of lunar ranging data within the SME framework using more than 26,000 normal points between 1969 and 2018. We also perform a jackknife analysis in order to provide realistic estimates of the systematic uncertainties. No deviation from Lorentz/CPT symmetries is reported. In addition, when fitting simultaneously for the 15 canonical SME coefficients for Lorentz/CPT-violations, we improve up to three orders of magnitude previous post-fit constraints from radio pulsars.
\end{abstract}

\maketitle


Lorentz Symmetry (LS) is at the core of two pillars of modern physics: General Relativity (GR) and the Standard Model of particle physics. GR, the current paradigm for the gravitational interaction, describes accurately gravitational phenomena over a very large range of distance scales \cite{will:2014la}. On the other hand, the Standard Model of particle physics provides an astonishingly accurate description of matter at the microscopic level and of non-gravitational interactions. While these two pillars of modern physics are known to be extremely successful, it is commonly admitted that they are not the ultimate description of Nature but rather some effective theories valid only in the low-energy limit. This assumption is motivated by the construction of a quantum theory of gravitation that would unify all the fundamental interactions; such a theory has not been successfully developed so far. In addition, observations requiring the introduction of Dark Matter and/or Dark Energy cannot be explained using GR and the Standard Model. Therefore, in the last decades, there have been an increasing interest to experimentally search for deviations from the fundamental principles of GR and the Standard Model in order to constrain possible new \emph{scenarii}. 

LS being at the core of both GR and the Standard Model, testing it is a way to probe a very large class of theoretical extensions of both. In order to search for a breaking of LS in an agnostic way, an effective field theory named the Standard-Model Extension (SME) has been developed \cite{colladay:1997vn,*colladay:1998ys,kostelecky:2004fk} to systematically consider all possible violations of LS including violations due to CPT-breaking. It contains GR, the Standard Model and all possible Lorentz/CPT-violating terms that can be constructed at the level of the Lagrangian, introducing a large numbers of new coefficients that can be constrained experimentally \cite{kostelecky:2011ly}. In the pure gravitational sector, the SME Lagrangian extends the standard Einstein-Hilbert action by including Lorentz/CPT-violating terms constructed by contracting new fields with some operators built from curvature tensors with increasing mass dimension \cite{bailey:2006uq,bailey:2015fk,kostelecky:2016nx,bailey:2017aa}.

The lower mass dimension term, the $d=4$ term, known as the minimal SME, breaks the LS but preserves the CPT symmetry. This term has been widely studied \cite{bailey:2006uq} and the related SME coefficients ($\bar s^{\mu\nu}$) have been constrained by various measurements: Lunar Laser Ranging (LLR) \cite{battat:2007uq,bourgoin:2016yu,*bourgoin:2017aa}, planetary ephemerides \cite{iorio:2012zr,*hees:2015sf}, pulsars timing \cite{shao:2014qd,*shao:2014rc,shao:2018ab}, Gravity Probe B \cite{bailey:2013kq}, Very Long Baseline Interferometry \cite{le-poncin-lafitte:2016yq}, gravimeters \cite{muller:2008kx,*chung:2009uq,*shao:2018aa,*flowers:2017aa}, gravitational waves \cite{kostelecky:2016nx,abbott:2017aa} and cosmic rays \cite{kostelecky:2015db} (see \cite{hees:2016aa,tasson:2016fk} for a review).

The next-to-leading term in the SME action, the $d=5$ term has recently been studied in \cite{bailey:2017aa,Kostelecky:2020hbb} and breaks both Lorentz and CPT symmetries leading to new phenomenological signatures. In particular, it leads to a modification of the equations of motion for the 2-body problem that depends on the relative velocity of the bodies. Interestingly, short-range laboratory experiments \cite{long:2015kx,*bailey:2015fk,*shao:2016aa,2017PhLB..766..137K}, that are very powerful to constrain higher-dimensional operators (with $d>5$), are mainly insensitive to these new velocity-dependent Lorentz/CPT-breaking terms. Instead pulsars timing turn out to be among the most sensitive probes \cite{bailey:2017aa} and has already provided first constraints on these Lorentz/CPT-violating coefficients \cite{shao:2018ab}. 

The first $d=5$ term in the nonminimal SME expansion of the gravity sector can be written as a quadratic effective action which reads as~\cite{bailey:2015fk,kostelecky:2016nx,bailey:2017aa}
\begin{equation}
    \mathcal L^{(5)}=-\frac{c^4}{128\pi G}h_{\mu\nu}q^{\mu\rho\alpha\nu\beta\sigma\gamma}\partial_\beta R_{\rho\alpha\sigma\gamma}
\end{equation}
with $c$ the speed of light in a vacuum, $G$ the Newton gravitational constant, $R_{\alpha\beta\gamma\delta}$ the linearized Riemann tensor built from the space-time metric $g_{\mu\nu}=\eta_{\mu\nu}+h_{\mu\nu}$ with $\eta_{\mu\nu}$ the Minkowski metric. The Lorentz/CPT-breaking coefficients $q^{\mu\rho\alpha\nu\beta\sigma\gamma}$ have length dimension and only 60 of them are independent due to symmetries of the Riemann tensor. Some combinations of the $q^{\mu\rho\alpha\nu\beta\sigma\gamma}$ occur frequently in the formalism so that it is appropriate to introduce the effective coefficients $K_{jklm}$ as follows \cite{bailey:2017aa,shao:2018ab}
\begin{align}
       K_{jklm} &= -\frac{1}{6} \left( q_{0jk0l0m} + q_{n0knljm} +
    q_{njknl0m} +  \right. \nonumber \\ &\hspace{2cm} \left.\mbox{permutations} \right) \,,
\end{align}
where ``permutations'' mean all symmetric permutations in the last three indices $klm$.

It can be shown that the orbital dynamics at the linearized order actually depends on 15 independent combinations of the fundamental SME coefficients \cite{bailey:2017aa,shao:2018ab}. The expressions of these 15 canonical coefficients in terms of the $q^{\mu\rho\alpha\nu\beta\sigma\gamma}$  and in terms of all $K_{jklm}$ are given in Tabs.~\ref{tab:K} and \ref{tab:Kmatrix} (from the appendix), respectively. The contribution to the 2-body equations of motion due to the mass dimension 5 operators for Lorentz/CPT-violations eventually reads
\begin{align}
    \left[\frac{\mathrm d^2 r^j}{\mathrm dt^2}\right]_{d=5} = &\frac{GM v^k}{c\,r^3}\Big(15 n^ln^mn^nn_{[j}K_{k]lmn}-3K_{[jk]ll}\nonumber \\
    &+9n^ln^mK_{[jk]lm}-9n_{[j}K_{k]ll}n^m\Big)\label{eq:accKSME}
\end{align}
with $r=|\bm r|$, where $\bm r=\bm r_2-\bm r_1$ is the relative position of the two bodies, $\bm n=\bm r/r$, and $\bm v=\bm v_2-\bm v_1$ their relative velocity \cite{bailey:2017aa}. The total mass is $M=m_1+m_2$ and $A_{[ij]}=\frac{1}{2}(A_{ij}-A_{ji})$.

The deviation from GR is non-static, proportional to the velocity and inversely proportional to the cubic distance between the two bodies. This term and the associated phenomenology is therefore extremely different from all GR corrections, PPN deviations \cite{will:2014la}, violations of the universality of free fall \cite{nordtvedt:1968kx,*nordtvedt:1968uq,*nordtvedt:1968ys,*nordtvedt:1968zm} and others LS-violating terms with mass dimension 4 that have already been considered in previous LLR analyzes \cite{bourgoin:2016yu}. In this paper, we use 50 years of LLR measurements in order to estimate the $d=5$ SME gravity coefficients.

\begin{table*}
\caption{\label{tab:K} Definition and estimates of the 15 canonical independent coefficients. Estimates are derived from a global LLR data analysis. Realistic estimate of each canonical SME coefficient $x_i$ is reported such as $x_i\pm\sigma_{\mathrm{stat}}(x_i)\pm\sigma_{\mathrm{syst}}(x_i)$.}
\begin{ruledtabular}
\begin{tabular}{l c c}
Canonical & Definition & Value and uncertainties $(\mathrm{m})$\\
\hline\vspace{-0.3cm}\\
$K_\mathrm{XXXY}$ & $\frac{1}{3}\left(-q^\mathrm{TXYTXTX}+q^\mathrm{TXYXYXY}+q^\mathrm{TXYXZXZ}-q^\mathrm{XYZXZXT}\right)$ & $(+0.7\pm0.4\pm2.9)\times10^3$\\
$K_\mathrm{XXXZ}$ & $\frac{1}{3}\left(q^\mathrm{TXYXYXZ}-q^\mathrm{TXZTXTX}+q^\mathrm{TXZXZXZ}+q^\mathrm{XYZXYXT}\right)$ & $(+0.8\pm0.9\pm5.9)\times10^3$\\
$K_\mathrm{XXYY}$ & $\frac{1}{3}\left(-2q^\mathrm{TXYTXTY}+2q^\mathrm{TXYXZYZ}+q^\mathrm{XYZXYZT}-2q^\mathrm{XYZXZYT}\right)$ & $(-0.4\pm1.3\pm8.4)\times10^3$\\
$K_\mathrm{XXYZ}$ & $\frac{1}{6}\left(-2q^\mathrm{TXYTXTZ}-2q^\mathrm{TXYXYYZ}-2q^\mathrm{TXZTXTY}+2q^\mathrm{TXZXZYZ}+q^\mathrm{XYZXYYT}-q^\mathrm{XYZXZZT}\right)$ & $(+0.5\pm0.2\pm1.6)\times10^4$\\
$K_\mathrm{XXZZ}$ & $\frac{1}{3}\left(-2q^\mathrm{TXYXZYZ}-2q^\mathrm{TXZTXTZ}+2q^\mathrm{XYZXYZT}-q^\mathrm{XYZXZYT}\right)$ & $(-1.9\pm0.6\pm4.1)\times10^4$\\
$K_\mathrm{XYYY}$ & $-q^{\rm TXYTYTY} +q^{\rm TXYXYXY} + q^{\rm TXYYZYZ} - q^{\rm XYZYZYT}$ & $(-0.7\pm0.3\pm1.2)\times10^4$\\
$K_\mathrm{XYYZ}$ & $\frac{1}{3} \left( -2 q^{\rm TXYTYTZ} +3 q^{\rm TXYXYXZ} -q^{\rm TXZTYTY} + q^{\rm TXZYZYZ} -q^{\rm XYZYZZT} \right)$ & $(+4.6\pm1.6\pm6.9)\times10^3$\\
$K_\mathrm{XYZZ}$ & $\frac{1}{3} \left( -q^{\rm TXYTZTZ} + 3 q^{\rm TXYXZXZ} + q^{\rm TXYYZYZ} - 2 q^{\rm TXZTYTZ} - q^{\rm XYZYZYT} \right)$ & $(-0.2\pm0.8\pm4.1)\times10^3$\\
$K_\mathrm{XZZZ}$ & $-q^{\rm TXZTZTZ} + q^{\rm TXZXZXZ} + q^{\rm TXZYZYZ}- q^{\rm XYZYZZT}$ & $(+1.2\pm0.3\pm1.3)\times10^4$\\
$K_\mathrm{YXXZ}$ & $\frac{1}{3} \left( 3 q^{\rm TXYTXTZ} + 3 q^{\rm TXYXYYZ} - q^{\rm TXZTXTY} + q^{\rm TXZXZYZ} + q^{\rm XYZXZZT} \right)$ & $(+0.1\pm0.3\pm2.3)\times10^4$\\
$K_\mathrm{YXYZ}$ & $\frac{1}{6} \left( 4 q^{\rm TXYTYTZ} -2 q^{\rm TXYXYXZ} - 2 q^{\rm TXZTYTY} + 2 q^{\rm TXZYZYZ} + q^{\rm XYZXYXT} + q^{\rm XYZYZZT} \right)$ & $(-4.7\pm0.8\pm4.0)\times10^3$\\
$K_\mathrm{YXZZ}$ & $\frac{1}{3} \left( 3 q^{\rm TXYTZTZ} - q^{\rm TXYXZXZ} - 3 q^{\rm TXYYZYZ} - 2 q^{\rm TXZTYTZ} + q^{\rm XYZXZXT} \right)$ & $(-1.6\pm0.5\pm2.4)\times10^3$\\
$K_\mathrm{YYYZ}$ & $\frac{1}{3} \left( q^{\rm TXYXYYZ} - q^{\rm TXZTYTY} + q^{\rm TYZYZYZ} + q^{\rm XYZXYYT} \right)$ & $(+0.9\pm0.3\pm1.8)\times10^4$\\
$K_\mathrm{YYZZ}$ & $\frac{1}{3}\left( 2 q^{\rm TXYXZYZ} - 2 q^{\rm TXZTYTZ} + q^{\rm XYZXYZT} +q^{\rm XYZXZYT} \right)$ & $(-1.5\pm0.5\pm3.4)\times10^4$\\
$K_\mathrm{YZZZ}$ & $-q^{\rm TXZTZTZ} + q^{\rm TXZXZYZ} + q^{\rm TYZYZYZ} + q^{\rm XYZXZZT}$ & $(-1.2\pm0.8\pm5.1)\times10^4$\\
\end{tabular}
\end{ruledtabular}
\end{table*}

LLR observations consist of high-precision measurements of the two-way light travel-time of short laser pulses between a LLR station on Earth and a corner cube retroreflector on the surface of the Moon. The monitoring of this travel-time has allowed scientists to study the Earth-Moon system and in particular to infer Moon's internal properties \cite{2001JGR...10627933W,doi:10.1029/2019GL082677} and to test the gravitational interaction by testing the universality of free fall \cite{2009IJMPD..18.1129W} and by performing several tests of LS \cite{battat:2007uq,bourgoin:2016yu,*bourgoin:2017aa} (see \cite{2019JGeod..93.2195M} for a review). Currently, the LLR dataset contains more than 26,000 normal points acquired by 6 LLR stations (McDonald Observatory in Texas, Observatoire de la C\^ote d'Azur in France, Haleakala Observatory in Hawaii, Apache-Point Observatory in New Mexico, Matera in Italy, and Wettzell in Germany) using 5 retroreflectors (Apollo XI, XIV, XV, and Lunokhod 1, 2). The data consists of normal points that combine time series of measured light travel time of photons averaged over several minutes in order to achieve a higher signal-to-noise ratio measurement of the lunar range at some characteristic epoch. Each normal point is thus characterized by one emission time and one time delay (together with some additional observational parameters such like the laser wavelength or the temperature and pressure at the site of observation). In 2015, a significant upgrade was made by the French LLR staff observers who demonstrated the efficiency of conducting observations at infrared (IR) wavelength \cite{2017A&A...602A..90C}. Owing notably to a better atmospheric transmission than the green wavelength, the IR wavelength allows one to increase the station efficiency and to improve the temporal homogeneity of the LLR observations over a synodic month \cite{doi:10.1029/2019EA000785}. The German LLR station, which has been running since 2018, also makes use of the IR wavelength. Nowadays, more than fifty years after his first detection that occurred on August 1969 and thanks to the common effort carried out by the LLR staff observers all around the world, LLR is still one of the most precise techniques for testing GR \cite{2019JGeod..93.2195M}.

The Lorentz/CPT-violating terms in \eqref{eq:accKSME} and the corresponding partial derivatives have been added to the set of integrated equations of motion within ELPN (Eph\'em\'eride Lunaire Parisienne Num\'erique) software \cite{bourgoin:2016yu,bourgoin:2017aa}. ELPN is a numerical ephemeris being fully relativistic at the first post-Newtonian order. It solves for a state vector which includes the orbital dynamics of the main solar system bodies, the rotational motion of the Moon, and the evolution of the difference between the terrestrial time and the barycentric dynamical time. The partial derivatives of the state vector with respect to initial conditions and physical parameters are integrated as well. The numerical modeling includes all contributions producing theoretical effects larger than the millimeter level such as the relativistic point-mass interactions between the relevant bodies, the Newtonian accelerations due to gravity field inhomogeneities of Earth and Moon, or the secular tidal acceleration of the orbit of the Moon. The numerical integration provides the position, velocity and orientation of the Moon which are then transformed into an estimation of the LLR normal points following the recommendations of the International Earth Rotation and Reference Systems Service \cite{2010ITN....36....1P}. The residuals are deduced by comparing the theoretical estimations with the measurements and are finally minimized with a standard weighted least-squares fit.

The \emph{modus operandi} for estimating the $d=5$ SME gravity coefficients proceeds as follows. First, a solution in pure GR is built by estimating 68 parameters which include: the lunar positions of the retroreflectors, the geocentric positions of the Apache-point and Haleakala stations (these two stations are currently not defined in the International Terrestrial Reference Frame), one light-time correction (offset) per instrument, the initial conditions for the orbit of the Moon and rotation (mantle and core), the masses of the Moon and the Earth-Moon barycenter, the Earth and Moon Love number $k_2$, the Earth rotational time-lag for semi-diurnal deformation, the Moon time-lag for solid-body tides, the Moon undistorted principal moment of inertia, the friction coefficient between the core and mantle of the Moon, and finally degree 2 and 3 mass multipole moments of the Moon. Let us emphasize that the Moon degree 2 and 3 gravity coefficients are not fixed to values determined from GRAIL (Gravity Recovery and Interior Laboratory) mission \cite{2014GeoRL..41.3382L,2014GeoRL..41.1452K}. There are two main reasons for this. First, as pointed out by \cite{2019AGUFM.G31B0649V,2019pavlov}, imposing the GRAIL gravity field during LLR analysis results in unexpected signatures in the LLR post-fit residuals suggesting that GRAIL and LLR dynamical models might be inconsistent. Nevertheless, these signatures can be absorbed by estimating the lunar degree 3 gravity coefficients which, consequently, are not consistent with GRAIL's anymore. Secondly, as it can be seen \emph{a posteriori}, the Lorentz/CPT-violating fields are mostly uncorrelated with Newtonian parameters except for the lunar degree 2 gravity coefficients. Therefore, estimating degree 2 allows us to ensure that a Lorentz/CPT-violating signal was not absorbed while determining the same degree in the GR framework from GRAIL. Unlike degree 3, the values of $J_2$ and $C_{22}$ determined from LLR are always fully compatible with GRAIL's estimates. The ELPN post-fit residuals determined in pure GR are reported in Tab.~\ref{tab:res}. 

\begin{table}[b]
  \caption{ELPN (in pure GR) post-fit residuals per LLR station and instrument. The mean and the standard deviation of the residuals are denoted by $\mu$ and $\sigma$, respectively. For each station/instrument, $N$ is the number of available observations and $N_r$ the number of rejected observations ($>3\sigma$).}
  \label{tab:res}
  \begin{tabularx}{\columnwidth}{l >{\centering\arraybackslash}X c c c c}
    \toprule
    Station (Instrument)  & Period   & $N$ & $N_r$ &  $\mu$ ($\mathrm{cm}$) & $\sigma$ ($\mathrm{cm}$) \\
    \hline
    McDonald ($2.7$-$\mathrm{m}$)     & 1969-1985 & 3604  & 92  & 14.0  & 34.7 \\
    McDonald (MLRS1)      & 1983-1988 & 631   & 74  & 7.3   & 29.3 \\
    McDonald (MLRS2)      & 1988-2015 & 3670  & 467 & -1.0  & 5.5  \\
    Grasse (Rubis)        & 1984-1986 & 1188  & 21  & 4.5   & 16.0 \\
    Grasse (Yag)          & 1987-2005 & 8324  & 51  & 0.0   & 4.1  \\
    Grasse (MeO green)    & 2009-2018 & 1937  & 23  & 0.2   & 1.8  \\
    Grasse (MeO IR)       & 2015-2018 & 3837  & 25  & -0.2  & 1.7  \\
    Haleakala             & 1984-1990 & 770   & 23  & -2.8  & 8.1 \\
    Matera                & 2003-2018 & 224   & 15  & -0.4  & 4.7  \\
    Apache-point (P1)     & 2006-2010 & 941   & 2   & 0.9   & 2.2  \\
    Apache-point (P2)     & 2010-2012 & 513   & 15  & 0.9   & 2.9  \\
    Apache-point (P3)     & 2012-2013 & 360   & 9   & 0.7   & 2.3  \\
    Apache-point (P4)     & 2013-2016 & 834   & 7   & 1.0   & 1.7  \\
    Wetzell               & 2018-2018 & 22    & 0   & 1.7   & 1.2 \\
    \botrule
  \end{tabularx}
\end{table}

\begin{table*}[ht!]
  \caption{Values of $\mathbf V^T$'s components from LLR global analysis. Values larger than $0.1$ have been highlighted for readability.}
  \label{tab:combi}
  \begin{tabularx}{\textwidth}{lccccccccccccccc}
    \toprule
    & $K_\mathrm{XXXY}$ & $K_\mathrm{XXXZ}$ & $K_\mathrm{XXYY}$ & $K_\mathrm{XXYZ}$ & $K_\mathrm{XXZZ}$ & $K_\mathrm{XYYY}$ & $K_{\rm XYYZ}$ & $K_{\rm XYZZ}$ & $K_{\rm XZZZ}$ & $K_{\rm YXXZ}$ & $K_{\rm YXYZ}$ & $K_{\rm YXZZ}$ & $K_{\rm YYYZ}$ & $K_{\rm YYZZ}$ & $K_{\rm YZZZ}$ \\
    \hline
    $c_1$ & \ 0.00 & \ 0.01 & \ 0.09 & \textbf{-0.21} & \textbf{\ 0.50} & \ 0.01 & -0.01 & \ 0.00 & -0.02 & \textbf{\ 0.25} & \ 0.01 & \ 0.00 & \textbf{-0.19} & \textbf{\ 0.43} & \textbf{\ 0.65} \\
    $c_2$ & \ 0.06 & \textbf{-0.13} & \ 0.07 & -0.03 & \textbf{-0.26} & \textbf{\ 0.49} & \textbf{-0.29} & \textbf{-0.11} & \textbf{-0.60} & \textbf{\ 0.29} & \textbf{\ 0.13} & \ 0.07 & \textbf{\ 0.18} & \textbf{-0.16} & \textbf{\ 0.20} \\
    $c_3$ & -0.02 & \ 0.04 & \textbf{\ 0.15} & -0.08 & \textbf{-0.45} & \textbf{-0.32} & \textbf{\ 0.19} & \ 0.09 & \textbf{\ 0.34} & \textbf{\ 0.45} & -0.02 & \ 0.00 & \textbf{\ 0.36} & \textbf{-0.19} & \textbf{\ 0.37} \\
    $c_4$ & \ 0.01 & -0.03 & \textbf{-0.28} & \textbf{\ 0.27} & \textbf{\ 0.22} & \ 0.04 & \ 0.00 & -0.09 & \textbf{\ 0.11} & \textbf{\ 0.56} & \textbf{-0.13} & -0.08 & \textbf{-0.46} & \textbf{-0.46} & -0.08 \\
    $c_5$ & \textbf{-0.17} & \textbf{\ 0.39} & \ 0.09 & -0.06 & \textbf{-0.11} & \textbf{\ 0.33} & \textbf{-0.13} & \textbf{-0.29} & \textbf{\ 0.15} & \textbf{-0.11} & \textbf{-0.61} & \textbf{-0.41} & \ 0.03 & \ 0.00 & \textbf{\ 0.11} \\
    $c_6$ & \textbf{\ 0.24} & \textbf{-0.57} & -0.07 & \textbf{\ 0.14} & \ 0.09 & \textbf{\ 0.21} & -0.01 & \textbf{-0.28} & \textbf{\ 0.31} & \textbf{\ 0.21} & \textbf{-0.17} & -0.05 & \textbf{\ 0.35} & \textbf{\ 0.35} & \textbf{-0.22} \\
    $c_7$ & \textbf{-0.18} & \textbf{\ 0.49} & \textbf{-0.21} & \textbf{\ 0.23} & \textbf{\ 0.13} & \textbf{-0.14} & -0.03 & \textbf{\ 0.12} & \textbf{-0.26} & \textbf{\ 0.39} & \ 0.03 & -0.07 & \textbf{\ 0.36} & \textbf{\ 0.37} & \textbf{-0.30} \\
    $c_8$ & \textbf{\ 0.33} & \textbf{\ 0.11} & \ 0.03 & \textbf{-0.14} & \ 0.06 & \textbf{-0.45} & \textbf{-0.75} & \textbf{-0.23} & \ 0.06 & \ 0.01 & -0.08 & \textbf{\ 0.17} & \ 0.04 & -0.07 & -0.04 \\
    $c_9$ & -0.08 & \ 0.01 & \textbf{-0.78} & \textbf{-0.37} & \textbf{-0.27} & \textbf{\ 0.17} & \textbf{-0.13} & \textbf{\ 0.15} & \textbf{\ 0.18} & -0.02 & -0.02 & \textbf{\ 0.18} & -0.07 & \textbf{\ 0.14} & \ 0.08 \\
    $c_{10}$ & \textbf{\ 0.10} & \textbf{-0.15} & \textbf{-0.25} & \textbf{-0.45} & \textbf{\ 0.11} & \textbf{-0.32} & \textbf{\ 0.27} & \textbf{-0.30} & \textbf{-0.35} & -0.01 & \ 0.02 & \textbf{-0.50} & \textbf{\ 0.13} & \textbf{-0.15} & -0.05 \\
    $c_{11}$ & \textbf{-0.14} & \ 0.06 & \textbf{\ 0.18} & \textbf{-0.51} & \textbf{\ 0.44} & \textbf{\ 0.27} & -0.01 & \textbf{\ 0.22} & \textbf{\ 0.17} & \textbf{\ 0.12} & -0.03 & \textbf{\ 0.20} & \textbf{\ 0.31} & \textbf{-0.34} & \textbf{-0.26} \\
    $c_{12}$ & \textbf{\ 0.66} & \textbf{\ 0.32} & -0.05 & -0.04 & -0.01 & \ 0.09 & \textbf{\ 0.43} & \textbf{-0.10} & \textbf{-0.14} & \ 0.00 & \textbf{-0.28} & \textbf{\ 0.41} & \ 0.00 & \ 0.00 & \ 0.00 \\
    $c_{13}$ & \textbf{-0.14} & \textbf{\ 0.21} & \textbf{-0.24} & \textbf{\ 0.25} & \textbf{\ 0.23} & \ 0.09 & \ 0.07 & \textbf{-0.53} & \textbf{\ 0.18} & -0.22 & \textbf{\ 0.39} & \textbf{\ 0.16} & \textbf{\ 0.32} & \textbf{-0.23} & \textbf{\ 0.26} \\
    $c_{14}$ & \textbf{-0.54} & \textbf{-0.18} & \ 0.06 & -0.09 & -0.07 & \textbf{-0.22} & \textbf{\ 0.13} & \textbf{-0.37} & \textbf{-0.21} & \ 0.07 & \textbf{-0.34} & \textbf{\ 0.52} & \textbf{-0.10} & \ 0.07 & -0.08 \\
    $c_{15}$ & \ 0.02 & \textbf{\ 0.23} & \textbf{\ 0.24} & \textbf{-0.31} & \textbf{-0.24} & \textbf{\ 0.15} & \ 0.07 & \textbf{-0.39} & \textbf{\ 0.20} & \textbf{\ 0.24} & \textbf{\ 0.46} & -0.04 & \textbf{-0.33} & \textbf{\ 0.24} & \textbf{-0.30} \\
    \botrule
  \end{tabularx}
\end{table*}

\begin{table}[hb!]
  \caption{Realistic estimates of linear combinations of SME coefficients (see Tab. \ref{tab:combi}) from a global LLR data analysis. Realistic estimate of each linear combination $c_i$ is reported such as $c_i\pm\sigma_{\mathrm{stat}}(c_i)\pm\sigma_{\mathrm{syst}}(c_i)$.}
  \label{tab:combvalue}
  \begin{tabularx}{\columnwidth}{>{\centering\arraybackslash}X>{\centering\arraybackslash}X}
    \toprule
    Linear combination & Value and uncertainties $(\mathrm{m})$ \\
    \hline
    $c_1$ & $(-2.7\pm1.1\pm7.8)\times10^4$\\
    $c_2$ & $(-0.6\pm0.4\pm1.4)\times10^4$\\
    $c_3$ & $(+1.8\pm0.4\pm2.7)\times10^4$\\
    $c_4$ & $(+3.4\pm1.2\pm5.9)\times10^3$\\
    $c_5$ & $(+3.6\pm1.2\pm4.6)\times10^3$\\
    $c_6$ & $(+2.4\pm0.7\pm8.7)\times10^3$\\
    $c_7$ & $(-2.0\pm0.7\pm2.9)\times10^3$\\
    $c_8$ & $(+0.9\pm0.2\pm1.6)\times10^3$\\
    $c_9$ & $(-2.0\pm0.8\pm2.1)\times10^2$\\
    $c_{10}$ & $(-3.5\pm1.0\pm5.6)\times10^2$\\
    $c_{11}$ & $(-1.8\pm0.9\pm5.0)\times10^2$\\
    $c_{12}$ & $(+0.1\pm0.2\pm2.0)\times10^3$\\
    $c_{13}$ & $(+0.4\pm0.1\pm1.5)\times10^2$\\
    $c_{14}$ & $(-1.0\pm0.4\pm3.9)\times10^2$\\
    $c_{15}$ & $(-0.3\pm0.1\pm1.0)\times10^2$\\
    \botrule
  \end{tabularx}
\end{table}

After convergence is achieved, the 15 SME gravity canonical coefficients are fitted simultaneously with the 68 Newtonian parameters. Two remarks need to be mentioned at this stage: (i) the obtained uncertainties (i.e. the covariance matrix) are only statistical uncertainties and do not contain any estimate of systematics and (ii) the estimates of the 15 canonical SME coefficients are highly correlated.

The statistical uncertainty estimated from a least-squares is not sufficient to provide ``realistic'' parameter uncertainties. Indeed, it is well known that the standard errors derived from least-squares analysis are prone to be too optimistic \cite{1992nrfa.book.....P,1996PhRvD..53.6730W}. As a matter of fact, parameter uncertainties are also affected by systematic errors which are not included in a standard least-squares. Therefore, in order to provide realistic parameter estimates, we look for systematics by using a jackknife resampling method \cite{1993stp..book.....L,bourgoin:2016yu,bourgoin:2017aa}. The method proceeds as follows. First, we build $n_{\mathrm{s}}$ subsets of data $\mathcal D_k$ with $k=1,\ldots,n_{\mathrm{s}}$ and $n_{\mathrm{s}}=6$ where $\mathcal D_k$ contains all the LLR observation except the ones from station $k$. For each subset $\mathcal D_k$, we perform a fit as described previously and we denote $x^{(k)}$ the estimated value of a certain parameter $x$. Then, considering that each subset provides an independent determination of $x$, an estimation of the variance of the systematic error due to stations is given by~\cite{1993stp..book.....L}
\begin{equation}
  \sigma_{\mathrm{s}}^2(x)=\frac{n_{\mathrm{s}}-1}{n_{\mathrm{s}}}\sum_{k=1}^{n_{\mathrm{s}}}\left[ x^{(k)}-\frac{1}{n_{\mathrm{s}}}\sum_{l=1}^{n_{\mathrm{s}}}x^{(l)}\right]^2\text{.}
\end{equation}
The exact same methodology is applied by building 5 additional subsets of data by lunar retroreflectors ($n_{\mathrm{r}}=5$), and by determining $\sigma_{\mathrm{r}}^2(x)$ the estimate of the systematic uncertainty due to retroreflectors. Finally, the total systematic variance of $x$ is $\sigma_{\mathrm{syst}}^2(x)=\sigma_{\mathrm{r}}^2(x)+\sigma_{\mathrm{s}}^2(x)$ such that the realistic variance is eventually given by $\sigma_{\mathrm{real}}^2(x)=\sigma_{\mathrm{stat}}^2(x)+\sigma^2_{\mathrm{syst}}(x)$, where $\sigma_{\mathrm{stat}}(x)$ is the statistical uncertainty of $x$. 

\begin{figure*}[hb!]
  \begin{center}
    \includegraphics[scale=0.388]{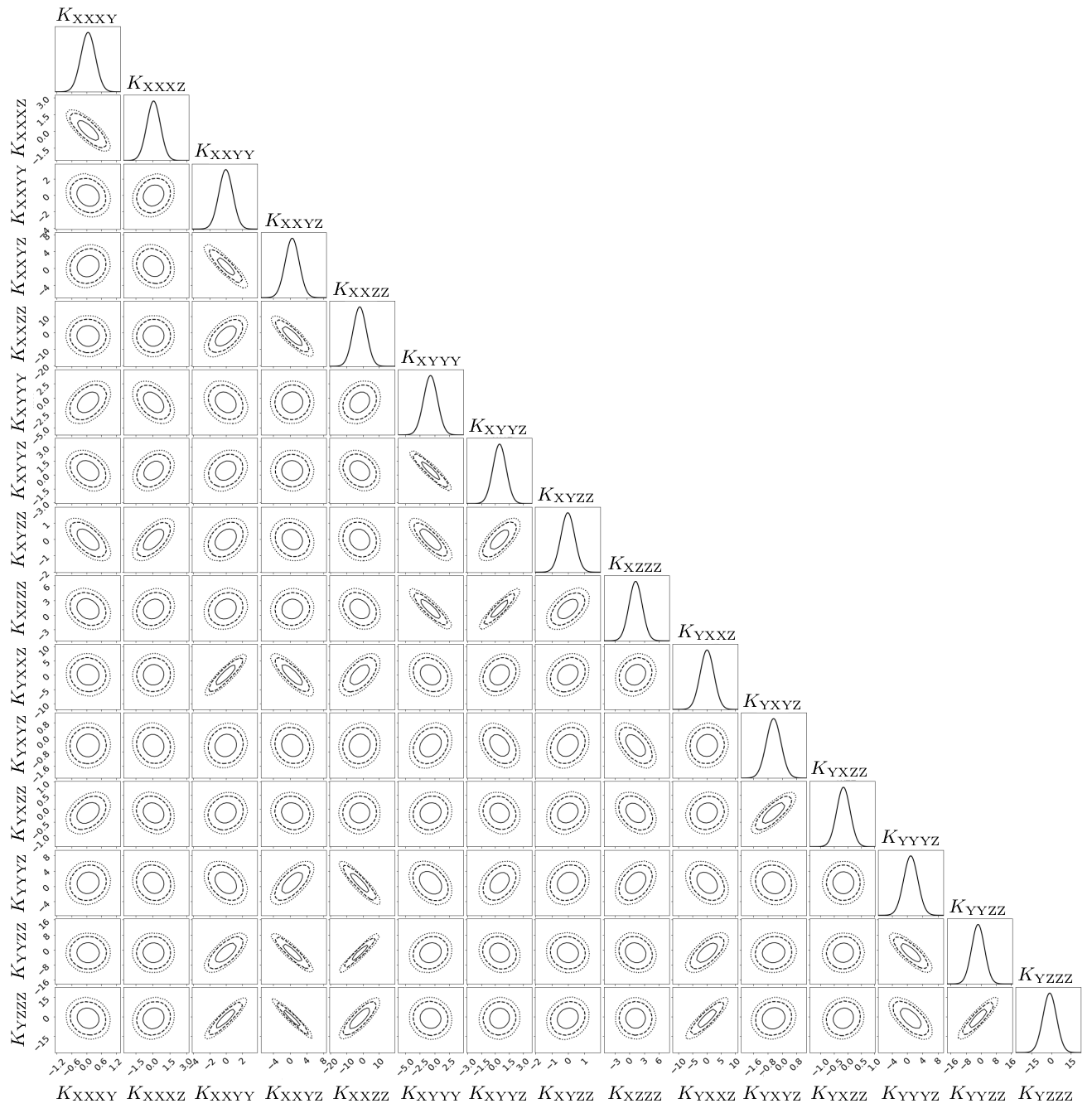}
  \end{center}
  \vspace{-0.5cm}
  \caption{Marginal 2D contours and marginal 1D posterior distribution of the set of 15 independent canonical SME coefficients from a LLR global LLR data analysis. Contours show the 68\%, 90\%, and 95\% Bayesian confidence levels. The canonical SME coefficients are expressed in $10^4\ \mathrm{m}$.}
  \label{fig_corner}
\end{figure*}

The estimates of the 15 canonical SME coefficients and their corresponding marginalized uncertainties are presented in Tab.~\ref{tab:K}. No deviation from GR is reported at 68\% Bayesian confidence level. The full 2-D posterior is presented in Fig.~\ref{fig_corner} from the appendix. This figure and estimates in Tab. \ref{tab:K} can be compared to Fig. 2 and Tab. VI of Shao and Bailey \cite{shao:2018ab}, respectively. The estimates of the 15 canonical SME coefficients from LLR global data analysis improve by 2 to 3 (see $K_\mathrm{XXXY}$ and $K_\mathrm{YXZZ}$) orders of magnitude previous constraints from radio pulsars. This improvement is mainly due to the fact that the huge number of LLR data and the timespan lead to a better decorrelation of the various SME coefficients while the pulsars analysis from \cite{shao:2018ab} relies only on 15 (although extremely powerful) estimates  of pulsars linear drift of the argument of periastron and of the projected semi-major axis. Finally, let us mention a major conceptual difference between the two data processing pipelines. We are analyzing LLR data directly in the SME framework by estimating simultaneously Newtonian parameters and SME coefficients. On the other hand, constraints inferred in \cite{shao:2018ab} rely on post-fit estimates based on a previous GR analysis which is known to be prone to return uncertainties that are too optimistic \cite{le-poncin-lafitte:2016yq}. In that sense global fit analysis are usually much more robust.

The SME coefficients estimated using LLR are still highly correlated (see Fig.~\ref{fig_corner} from the appendix). One can use a single value decomposition of the covariance matrix to set estimates on independent linear combinations of them. Let $\bm a$ be the set of the 15 canonical coefficients and $\bm c$ be the set of the 15 independent linear combinations of $\bm a$.  The independent linear combinations are thus given by $\bm c=\textbf V^T\cdot \bm a$, where the matrix $\textbf V$ is determined by diagonalizing the covariance matrix $\textbf C$ by performing a singular value decomposition \cite{1992nrfa.book.....P}, that is to say $\textbf C=\textbf V\cdot\textbf W\cdot\textbf V^T$. The matrix $\textbf V$ contains the eigenvectors of $\textbf C$ and the diagonal matrix $\textbf W$ contains the eigenvalues of $\textbf C$, hence the statistical uncertainty associated to the independent linear combinations $c_i$ is given by $\sigma_{\mathrm{stat}}^2(c_i)=W_{ii}$. The detailed linear combinations are presented in Tab.~\ref{tab:combi} and their estimations are given in Tab. \ref{tab:combvalue}. The advantage of using the linear combinations relies in the fact that their estimations are independent. Uncertainty estimates in Tab. \ref{tab:combvalue} range between $79\ \mathrm{km}$ to $110\ \mathrm{m}$, in agreement with \cite{bailey:2017aa}. 


In conclusion, in this Letter, we use 50 years of LLR data to search for a velocity-dependent Lorentz/CPT-breaking signature. Such a phenomenology, induced by order 5 terms in the pure gravity SME Lagrangian \cite{bailey:2017aa}, is relatively novel and has only been considered in pulsars analysis \cite{shao:2018ab}. Fitting simultaneously the SME coefficients with all standard LLR parameters, we report no breaking of Lorentz/CPT symmetries and report realistic estimates on the 15 canonical coefficients at the level of $10^{3-4}$ m improving similar constraints from pulsars by up to 3 orders of magnitude. This improvement is mainly due to the large numbers of data used in our analysis which leads to smaller correlation between the various $K$ coefficients, while pulsar individual constraints are more stringent. This would suggest performing a combine fit using both LLR and pulsars data simultaneously.

\par {\it Acknowledgments.}
This work was supported by the Programme National GRAM of CNRS/INSU with INP and IN2P3 co-funded by CNES. Q.G.B. supported by NSF grant number 1806871.

\bibliography{SME_bib}

\appendix*

\section{Supplemental Material}

\begin{table*}[b!]
\caption{\label{tab:Kmatrix} Components $K_{jklm}$ in terms of the 15 independent canonical SME coefficients.}
\begin{ruledtabular}
\begin{tabularx}{\textwidth}{ll|ll|ll|ll} 
  $K_{jklm}$ & Canonical & $K_{jklm}$ & Canonical & $K_{jklm}$ & Canonical & $K_{jklm}$ & Canonical \\
  \hline
$K_\mathrm{XXXX}$ &  0                  & $K_\mathrm{XZXZ}$ & $K_\mathrm{XXZZ}$      &  $K_\mathrm{YYYY}$ & 0                                      &  $K_\mathrm{ZXZX}$ & $-K_\mathrm{XXZZ}$ \\                    
$K_\mathrm{XXXY}$ & $K_\mathrm{XXXY}$   & $K_\mathrm{XZYX}$ & $K_\mathrm{XXYZ}$      &  $K_\mathrm{YYYZ}$ & $K_\mathrm{YYYZ}$                      &  $K_\mathrm{ZXZY}$ & $-\frac{1}{2}K_\mathrm{XYZZ}-\frac{1}{2}K_\mathrm{YXZZ}$ \\
$K_\mathrm{XXXZ}$ & $K_\mathrm{XXXZ}$   & $K_\mathrm{XZYY}$ & $K_\mathrm{XYYZ}$      &  $K_\mathrm{YYZX}$ & $K_\mathrm{YXYZ}$                      &  $K_\mathrm{ZXZZ}$ & $-\frac{1}{3}K_\mathrm{XZZZ}$ \\
$K_\mathrm{XXYX}$ & $K_\mathrm{XXXY}$   & $K_\mathrm{XZYZ}$ & $K_\mathrm{XYZZ}$      &  $K_\mathrm{YYZY}$ & $K_\mathrm{YYYZ}$                      &  $K_\mathrm{ZYXX}$ & $-2K_\mathrm{XXYZ}-K_\mathrm{YXXZ}$ \\
$K_\mathrm{XXYY}$ & $K_\mathrm{XXYY}$   & $K_\mathrm{XZZX}$ & $K_\mathrm{XXZZ}$      &  $K_\mathrm{YYZZ}$ & $K_\mathrm{YYZZ}$                      &  $K_\mathrm{ZYXY}$ & $-K_\mathrm{XYYZ}-2K_\mathrm{YXYZ}$ \\
$K_\mathrm{XXYZ}$ & $K_\mathrm{XXYZ}$   & $K_\mathrm{XZZY}$ & $K_\mathrm{XYZZ}$      &  $K_\mathrm{YZXX}$ & $K_\mathrm{YXXZ}$                      &  $K_\mathrm{ZYXZ}$ & $-\frac{1}{2}K_\mathrm{XYZZ}-\frac{1}{2}K_\mathrm{YXZZ}$ \\
$K_\mathrm{XXZX}$ & $K_\mathrm{XXXZ}$   & $K_\mathrm{XZZZ}$ & $K_\mathrm{XZZZ}$      &  $K_\mathrm{YZXY}$ & $K_\mathrm{YXYZ}$                      &  $K_\mathrm{ZYYX}$ & $-K_\mathrm{XYYZ}-2K_\mathrm{YXYZ}$ \\
$K_\mathrm{XXZY}$ & $K_\mathrm{XXYZ}$   & $K_\mathrm{YXXX}$ & $-3K_\mathrm{XXXY}$    &  $K_\mathrm{YZXZ}$ & $K_\mathrm{YXZZ}$                      &  $K_\mathrm{ZYYY}$ & $-3K_\mathrm{YYYZ}$ \\
$K_\mathrm{XXZZ}$ & $K_\mathrm{XXYZ}$   & $K_\mathrm{YXXY}$ & $-K_\mathrm{XXYY}$     &  $K_\mathrm{YZYX}$ & $K_\mathrm{YXYZ}$                      &  $K_\mathrm{ZYYZ}$ & $-K_\mathrm{YYZZ}$ \\
$K_\mathrm{XYXX}$ & $K_\mathrm{XXXY}$   & $K_\mathrm{YXXZ}$ & $K_\mathrm{YXXZ}$      &  $K_\mathrm{YZYY}$ & $K_\mathrm{YYYZ}$                      &  $K_\mathrm{ZYZX}$ & $-\frac{1}{2}K_\mathrm{XYZZ}-\frac{1}{2}K_\mathrm{YXZZ}$ \\
$K_\mathrm{XYXY}$ & $K_\mathrm{XXYY}$   & $K_\mathrm{YXYX}$ & $-K_\mathrm{XXYY}$     &  $K_\mathrm{YZYZ}$ & $K_\mathrm{YYZZ}$                      &  $K_\mathrm{ZYZY}$ & $-K_\mathrm{YYZZ}$ \\
$K_\mathrm{XYXZ}$ & $K_\mathrm{XXYZ}$   & $K_\mathrm{YXYY}$ & $-\frac{1}{3}K_\mathrm{XYYY}$   &  $K_\mathrm{YZZX}$ & $K_\mathrm{YXZZ}$                      &  $K_\mathrm{ZYZZ}$ & $-\frac{1}{3}K_\mathrm{YZZZ}$ \\
$K_\mathrm{XYYX}$ & $K_\mathrm{XXYY}$   & $K_\mathrm{YXYZ}$ & $K_\mathrm{YXYZ}$      &  $K_\mathrm{YZZY}$ & $K_\mathrm{YYZZ}$                      &  $K_\mathrm{ZZXX}$ & $-K_\mathrm{XXZZ}$ \\
$K_\mathrm{XYYY}$ & $K_\mathrm{XYYY}$   & $K_\mathrm{YXZX}$ & $K_\mathrm{YXXZ}$      &  $K_\mathrm{YZZZ}$ & $K_\mathrm{YZZZ}$                      &  $K_\mathrm{ZZXY}$ & $-\frac{1}{2}K_\mathrm{XYZZ}-\frac{1}{2}K_\mathrm{YXZZ}$ \\
$K_\mathrm{XYYZ}$ & $K_\mathrm{XYYZ}$   & $K_\mathrm{YXZY}$ & $K_\mathrm{YXYZ}$      &  $K_\mathrm{ZXXX}$ & $-3K_\mathrm{XXXZ}$                    &  $K_\mathrm{ZZXZ}$ & $-\frac{1}{3}K_\mathrm{XZZZ}$ \\
$K_\mathrm{XYZX}$ & $K_\mathrm{XXYZ}$   & $K_\mathrm{YXZZ}$ & $K_\mathrm{YXZZ}$      &  $K_\mathrm{ZXXY}$ & $-2K_\mathrm{XXYZ}-K_\mathrm{YXXZ}$    &  $K_\mathrm{ZZYX}$ & $-\frac{1}{2}K_\mathrm{XYZZ}-\frac{1}{2}K_\mathrm{YXZZ}$ \\
$K_\mathrm{XYZY}$ & $K_\mathrm{XYYZ}$   & $K_\mathrm{YYXX}$ & $-K_\mathrm{XXYY}$     &  $K_\mathrm{ZXXZ}$ & $-K_\mathrm{XXZZ}$                     &  $K_\mathrm{ZZYY}$ & $-K_\mathrm{YYZZ}$ \\
$K_\mathrm{XYZZ}$ & $K_\mathrm{XYZZ}$   & $K_\mathrm{YYXY}$ & $-\frac{1}{3}K_\mathrm{XYYY}$   &  $K_\mathrm{ZXYX}$ & $-2K_\mathrm{XXYZ}-K_\mathrm{YXXZ}$    &  $K_\mathrm{ZZYZ}$ & $-\frac{1}{3}K_\mathrm{YZZZ}$ \\
$K_\mathrm{XZXX}$ & $K_\mathrm{XXXZ}$   & $K_\mathrm{YYXZ}$ & $K_\mathrm{YXYZ}$      &  $K_\mathrm{ZXYY}$ & $-2K_\mathrm{YXYZ}-K_\mathrm{XYYZ}$    &  $K_\mathrm{ZZZX}$ & $-\frac{1}{3}K_\mathrm{XZZZ}$ \\
$K_\mathrm{XZXY}$ & $K_\mathrm{XXYZ}$   & $K_\mathrm{YYYX}$ &$-\frac{1}{3}K_\mathrm{XYYY}$    &  $K_\mathrm{ZXYZ}$ &$-\frac{1}{2}K_\mathrm{XYZZ}-\frac{1}{2}K_\mathrm{YXZZ}$  &  $K_\mathrm{ZZZY}$ &$-\frac{1}{3}K_\mathrm{YZZZ}$   \\ 
$K_\mathrm{ZZZZ}$ &0 &&&&&
\end{tabularx}
\end{ruledtabular}
\end{table*}

\end{document}